\documentclass[11pt,onecolumn]{article}
\usepackage{float}
\usepackage{graphicx}
\usepackage{subfigure}
\usepackage{latexsym,amssymb}
\usepackage{amsmath, amsbsy}
\usepackage{amsopn, amstext}
\usepackage{psfrag, epsfig}
\usepackage{cite}
\usepackage{algorithm}
\usepackage{algpseudocode}
\usepackage{booktabs}
\usepackage{caption}
\usepackage[utf8]{inputenc}
\usepackage{multirow}
\usepackage{makecell}
\usepackage{color}
\usepackage{setspace}
\usepackage{arydshln}
\makeatletter
\newenvironment{breakablealgorithm}
  {
   \begin{center}
     \refstepcounter{algorithm}
     \hrule height.8pt depth0pt \kern2pt
     \renewcommand{\caption}[2][\relax]{
       {\raggedright\textbf{\ALG@name~\thealgorithm} ##2\par}%
       \ifx\relax##1\relax 
         \addcontentsline{loa}{algorithm}{\protect\numberline{\thealgorithm}##2}%
       \else 
         \addcontentsline{loa}{algorithm}{\protect\numberline{\thealgorithm}##1}%
       \fi
       \kern2pt\hrule\kern2pt
     }
  }{
     \kern2pt\hrule\relax
   \end{center}
  }
\makeatother

\newtheorem{theorem}{Theorem}[section]  
\newtheorem{lemma}[theorem]{Lemma}

\makeatletter \@addtoreset{equation}{section}\makeatother

\newcommand{\be}{\begin{equation}}
\newcommand{\ee}{\end{equation}}

\newcommand{\bea}{\begin{eqnarray}}
\newcommand{\eea}{\end{eqnarray}}

\def\ba{\begin{array}}
\def\ea{\end{array}}

\renewcommand\baselinestretch{1.2}

\def\i{{\bf i}}
\def\j{{\bf j}}
\def\k{{\bf k}}

\usepackage[symbol]{footmisc}

\begin{document}
\begin{center}
{\Large\bf A fragile zero-watermarking method based on dual quaternion matrix decomposition}
\vspace*{.5\baselineskip}

{{\sc Mingcui Zhang} \hskip .51cm and \hskip .51cm {\sc Zhigang Jia}\footnote{Corresponding author. {Email}: zhgjia@jsnu.edu.cn}} \vspace*{.5\baselineskip}

{\small
School of Mathematics and Statistics, Jiangsu Normal University, Xuzhou 221116, P.R. China
}
\end{center}

\begin{abstract}
Medical images play a crucial role in assisting diagnosis, remote consultation, and academic research. However, during the transmission and sharing process, they face serious risks of copyright ownership and content tampering. Therefore, protecting medical images is of great importance. As an effective means of image copyright protection, zero-watermarking technology focuses on constructing watermarks without modifying the original carrier by extracting its stable features, which provides an ideal approach for protecting medical images. This paper aims to propose a fragile zero-watermarking method based on dual quaternion matrix decomposition, which utilizes the operational relationship between the standard part and the dual part of dual quaternions to correlate the medical image with patient information image, and generates zero-watermarking information based on the characteristics of dual quaternion matrix decomposition, ultimately achieving copyright protection and content tampering detection for medical images.
\end{abstract}


\section{Introduction}

With the continuous development of digital technology, digitalization has deeply penetrated the medical field \cite{FuMengZhan}. Medical images, as the core part of medical diagnostic procedures, provides direct diagnostic basis for clinicians by presenting objective images of the internal structure and function of the human body \cite{Pareek}. Based on these images, on the one hand, doctors can accurately identify the location of the lesion, assess the nature and severity of the lesion, and complete the localization, qualitative and quantitative analysis of the disease. On the other hand, it makes remote diagnosis and expert consultation across different locations and institutions possible, allowing top medical resources to cover a wider area\cite{WeiZhangYang}. It has especially won precious diagnostic time for patients in remote areas and those with acute and severe conditions. This inevitably requires the transmission of medical images and patient information between medical systems. During the transmission process, these data may face security issues such as malicious theft and tampering. Therefore, ensuring the integrity and authenticity of medical image information is crucial.

Watermarking technology, as an effective means of digital rights management distinct from traditional encryption-based methods, not only ensures the confidentiality of digital information content but also addresses the copyright protection and content authentication issues of medical images\cite{WeiZhangYang}, thus attracting extensive attention from scholars. Traditional watermarking technology \cite{Van,ChenJiaPeng1,ChenJiaPeng2,ZhangDingLi} refers to the technique of embedding specific information in a digital medium in an invisible or nearly invisible form. This information can prove copyright ownership, track infringement, or provide other additional information without affecting the original media content. Although it can be used to address these risks, it usually requires modifying the original image data, which must be avoided in medical scenarios as any minor change in data features may lead to incorrect diagnoses \cite{Ali}. This has promoted the application of zero-watermarking technology in the protection of medical images.

Zero-watermarking \cite{WenSunWang} refers to the process of using the inherent and most significant features of the carrier to generate a unique identity identifier (zero-watermarking), then registering and certifying this identifier with an independent and secure third-party institution. This approach can ensure the validity of the source of the original medical image and that it belongs to the correct patient without damaging the original medical image. Many scholars have conducted research on it.
Xia et al. \cite{XiaWangWang} extended the integer-order radial harmonic Fourier moments (IoRHFMs) to fractional-order radial harmonic Fourier moments (FoRHFMs), and proposed a medical image zero-watermarking algorithm based on FoRHFMs to achieve lossless copyright protection for medical images.
Xiang et al. \cite{XiangLiuLi} proposed a zero-watermarking medical image protection scheme based on style features and residual networks, which could better resist geometric attacks and achieve copyright protection.
Nawaz et al. \cite{Nawaz} proposed a medical image zero-watermarking technique based on dual-tree complex wavelet transform-AlexNet and discrete cosine transform, which successfully overcame the challenges of protecting watermarks from both conventional and geometric attacks.

Most of the above methods used zero-watermarking to protect the copyright or privacy of medical images.  Ali et al. \cite{Ali} further considered combining the fragile nature of the watermark with zero-watermarking, conducting research on fragile zero-watermarking to achieve both privacy protection and content authentication simultaneously. They proposed a new hybrid fragile zero-watermarking method based on visual cryptography and chaotic randomness and developed a breast cancer detection system using a convolutional neural network to analyze the diagnostic results after suffering from malicious attacks and watermark insertion.

Inspired by this, we propose a fragile zero-watermarking method based on the decomposition of dual quaternion matrices. This method binds medical images with patient information to form a dual quaternion matrix, and uses dual quaternion operations to achieve overall processing of medical images and patient information. During the zero-watermarking generation stage, the patient information image (watermark image) is encrypted and protected using Arnold transformation, the features of the medical image (original carrier image) are extracted using the fast Fourier transform, the relationship between the real part and the dual part of the dual quaternion is used to associate the medical image with the patient information image, and the zero-watermarking information is generated based on the characteristics of the dual quaternion matrix decomposition. In the watermark verification stage, the patient information image (watermark image) is reconstructed by using the verified medical image (verified carrier image) and the zero-watermarking information, and various indicators are calculated to obtain the verification result. Through these two stages, the copyright protection and content tampering detection of medical images are ultimately achieved.

The remainder of the paper is organized as follows. In Section 2, the fragile zero-watermarking method is analyzed from three aspects: preprocessing of the original carrier image and watermark image, generation of zero-watermarking, and verification of watermark. In Section 3, the fragile zero-watermarking generation algorithm and watermark verification algorithm are proposed and validated through experiments. In Section 4, the full paper is summarized.

\section{Analysis of fragile zero-watermarking model}

Unlike traditional watermarking, zero-watermarking does not embed a watermark into the original carrier image, but rather generates zero-watermarking information by integrating the features of original carrier image with the watermark. The general process of the zero-watermarking model can be described as follows.

\begin{figure}[H]
\centering
\includegraphics[width=16cm]{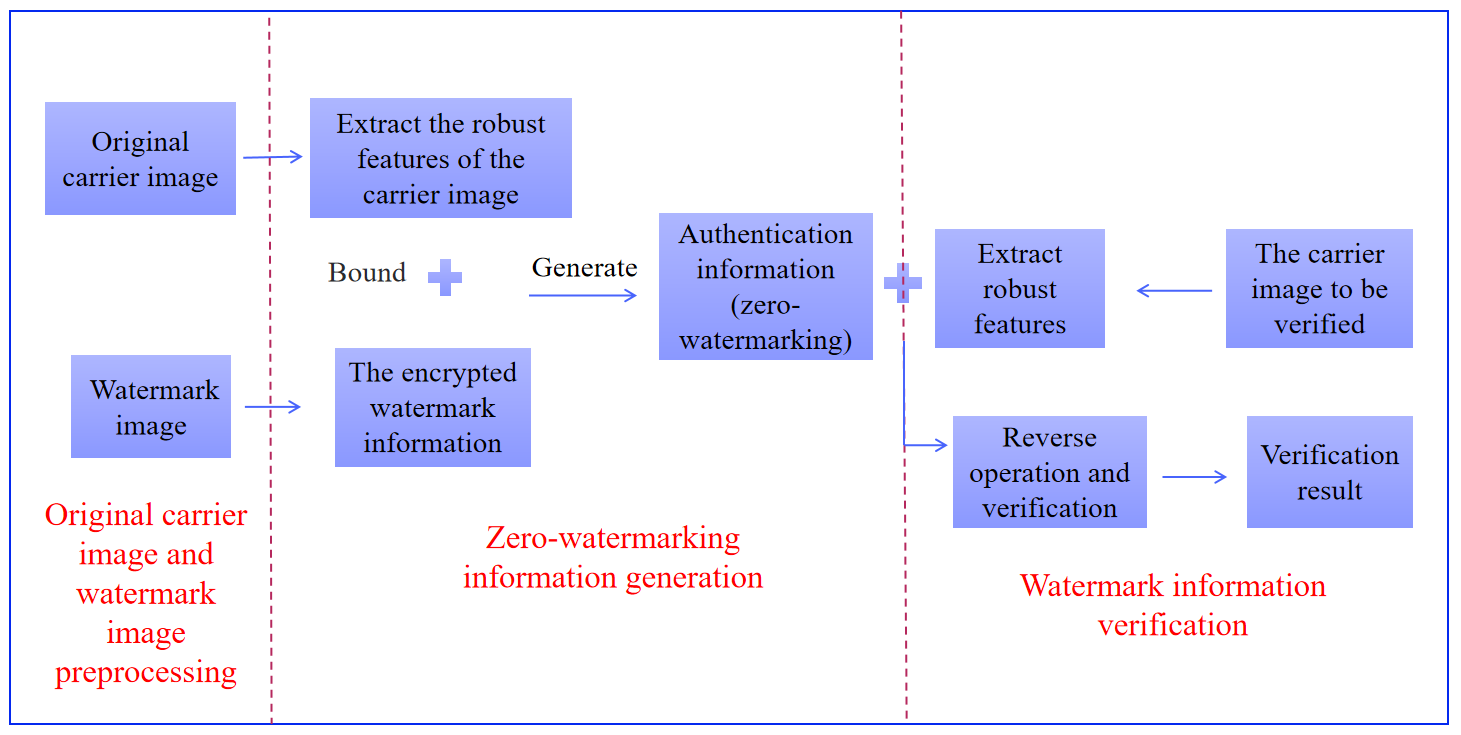}
\caption{Zero-watermarking process.}
\label{f2.1}
\end{figure}

Next, we will introduce the implementation of our algorithm in three stages: preprocessing of the original carrier image and watermark image, generation of zero-watermarking information, and verification of watermark information.

\subsection{Preprocessing of the original carrier image and the watermark image}

\subsubsection{Watermark image preprocessing process}

During the preprocessing stage of the watermark image, the Arnold scrambling is utilized to encrypt the watermark image. Arnold scrambling\cite{ChenQuanTay} is periodic and used to keep the watermarking secure. For a $N\times N$ image, let $(x,y)$ is the old pixel coordinates of the original image, $(\hat{x}, \hat{y})$ is the new pixel coordinates after iterative computation scrambling and $a$, $b$, $c$, $d$ are any positive integers, which satisfy $ad - bc = \pm 1$, then Arnold scrambling can be expressed as
\setlength{\arraycolsep}{5pt}
$$
\begin{bmatrix}\hat{x}\\\hat{y}\end{bmatrix}=\begin{bmatrix}a & b \\ c & d \end{bmatrix}\begin{bmatrix}x\\y\end{bmatrix}\mathrm{mod} N,
\begin{bmatrix}x\\y\end{bmatrix}=\begin{bmatrix}a & b \\ c & d \end{bmatrix}^{-1}\begin{bmatrix}\hat{x}\\\hat{y}\end{bmatrix}\mathrm{mod} N.
$$

\begin{figure}[H]
\centering
\includegraphics[width=10cm]{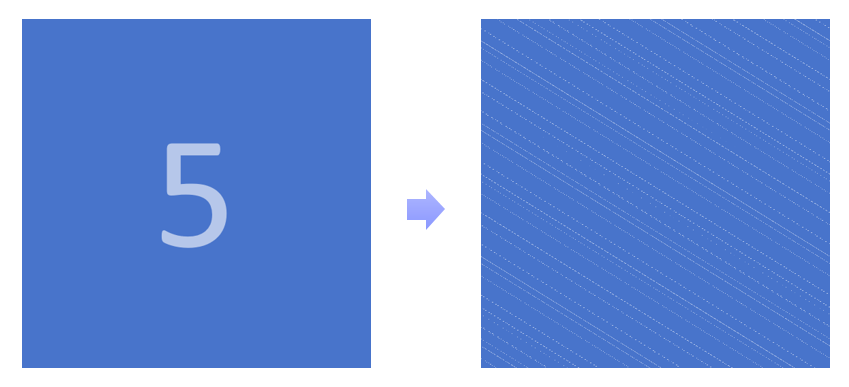}
\caption{Arnold scrambling process.}
\label{f2.2}
\end{figure}

\subsubsection{Original carrier image preprocessing process}

During the preprocessing stage of the original carrier image, the fast Fourier transform is used to convert the original carrier image from the spatial domain to the frequency domain, and its frequency domain features are extracted.

The fast Fourier Transform \cite{Cooley} was proposed by Cooley and Tukey in 1965, which reduced the computational complexity from $ O(N^2)$ to $O(N log N)$. We use the two-dimensional Fourier transform to preprocess the original carrier image, thereby obtaining high-quality images. Let $f(i,j)$ is the original matrix values, and $F(u,v)$ is the output matrix values. Suppose the image yields an $M \times N$ matrix. The forward transform is calculated using
$$
F(u,v)=\sum\limits_{x=0}^{M-1} \sum\limits_{y=0}^{N-1}f(x,y)exp \left[-2\pi j \left(\frac{ux}{M}+\frac{vy}{N}\right)\right],
$$
and the inverse transform is calculated using
$$
f(x,y)=\frac{1}{MN}\sum\limits_{x=0}^{M-1} \sum\limits_{y=0}^{N-1}F(u,v)exp \left[2\pi j \left(\frac{ux}{M}+\frac{vy}{N}\right)\right],
$$
where $u, v$ are frequency variables and $x, y$ are spatial variables.

%
%

\subsection{Zero-watermarking generation process}

During the zero-watermarking generation stage, the characteristics of the original carrier information and the encrypted watermark information are represented as a dual quaternion matrix. Then, the dual quaternion LU (DQLU) decomposition, QR (DQQR) decomposition, and singular value (DQSVD) decomposition are utilized to decompose the dual quaternion matrix respectively, and the partial information of the decomposed matrices is used as the zero-watermarking information. In this section, we will introduce the knowledge mentioned above. First, we will introduce the relevant knowledge of dual numbers and dual quaternions.

\subsubsection{Relevant knowledge}

Denote $\mathbb{R}$, $\mathbb{D}$, $\mathbb{Q}$ and $\mathbb{DQ}$ as the set of real numbers, dual numbers, quaternions and dual quaternions, respectively. $\mathbb{F}^{m}$ and $\mathbb{F}^{m\times n}$ are denoted the set of all $m\times 1$ vectors and $m\times n$ matrices on $\mathbb{F}$. $I_t$ represents $t \times t$ identity matrix, $O_{m \times n}$ represents $m \times n$ identity matrix.
A dual number \cite{QiLingYan} $d$ can be written as $d=d_s+d_i\epsilon$, where $d_s, d_i \in \mathbb{R}$ are the standard (real) part and the infinitesimal (dual) part of $d$, respectively, $\epsilon$ is the infinitesimal unit satisfying $\epsilon \neq 0, \epsilon^2=0$ and is commutative in multiplication with complex numbers. If $d_s\neq 0$, $d$ is appreciable, otherwise, $d$ is infinitesimal.

For two dual numbers $d=d_s+d_i\epsilon, b=b_s+b_i\epsilon \in \mathbb{D}$, the multiplication and division \cite{CuiQi} are defined as
$$
db=d_sb_s+(d_sb_i+d_ib_s)\epsilon,
$$
$$
\displaystyle\frac{d}{b}=\left\{\begin{aligned}\displaystyle\frac{d_s}{b_s}
+(\displaystyle\frac{d_i}{b_s}-\displaystyle\frac{d_s}{b_s}\displaystyle\frac{b_i}{b_s})\epsilon,\
\ b_s\neq 0, \\
   \displaystyle\frac{d_i}{b_i}+c\epsilon,  \ \ d_s,b_s= 0,\end{aligned}\right.
$$
where $c \in \mathbb{R}$ is any real number. The total order \cite{QiLingYan} $\leq$ can be defined as $d\leq b$, if either $d_s < b_s$, or $d_s = b_s$ and $d_i \leq b_i$. In particular, $d$ is positive, nonnegative, nonpositive or negative, if $d > 0$, $d \geq 0$, $d \leq 0$ or $d < 0$, respectively.
Suppose that $d=d_s+d_i\epsilon \in \mathbb{D}$, and $d_s>0$, the square root\cite{QiLingYan} is defined as
$$
\sqrt{d}=\sqrt{d_s}+\displaystyle\frac{d_i}{2\sqrt{d_s}}\epsilon,
$$
and $\sqrt{d}=0$, when $d=0$.

A dual number vector can be defined as $\mathbf{d}=\mathbf{d}_s+\mathbf{d}_i\epsilon$, where $\mathbf{d}_s, \mathbf{d}_i \in \mathbb{R}^n$. If $\mathbf{d}_s\neq \mathbf{0}$, then $\mathbf{d}$ is appreciable, and a dual number matrix $A=(a_{ij}) \in \mathbb{D}^{m\times n}$ can be denoted as $A=A_s+A_i\epsilon$, where $A_s, A_i\in \mathbb{R}^{m\times n}$ are the standard part and the infinitesimal part of $A$, respectively. The transpose can be expressed as $A^T=A_s^T+A_i^T\epsilon$. A dual number square matrix $A$ is referred to orthogonal if $A^TA=AA^T=I$; Symmetric if $A^T=A$.

%

Before introducing dual quaternions, we review the relevant definitions of quaternions \cite{WeiLiZhang}, a quaternion $g$ can be represented as $g=g_1+g_2\i+g_3\j+g_4\k, g_t\in \mathbb{R}, t=1,2,3,4$, where $\i^2=\j^2=\k^2=-1, \i\j=-\j\i=\k, \i\k=-\k\i=-\j, \j\k=-\k\j=\i$. The conjugate and magnitude of $g$ are $\overline{g}=g_1-g_2\i-g_3\j-g_4\k$ and $|g|=\sqrt{g_1^2+g_2^2+g_3^2+g_4^2}$, respectively, and the inverse of a nonzero quaternion $g$ is $g^{-1}=\frac{\overline{g}}{|g|^2}$. A $n \times n$ quaternion matrix can be defined as $A=A_1+A_2\i+A_3\j+A_4\k=A_{\alpha}+A_{\beta}\j, A_t\in \mathbb{R}^{n \times n}, t=1,2,3,4, A_{\Gamma}\in \mathbb{C}^{n \times n}, \Gamma=\alpha, \beta$, the conjugate, transpose, conjugate transpose, Frobenius norm and $q$-determinant \cite{ZhangFZ} of $A$ are $\overline{A}=\overline{A}_1+\overline{A}_2\i+\overline{A}_3\j+\overline{A}_4\k$, $A^T=A_1^T+A_2^T\i+A_3^T\j+A_4^T\k$, $A^H=A_1^H+A_2^H\i+A_3^H\j+A_4^H\k$, $\|A\|_F=\sqrt{A_1^2+A_2^2+A_3^2+A_4^2}$ and $|A|_q=\left|\begin{bmatrix}A_1 & A_2\\-\overline{A_2}&\overline{A_1}\end{bmatrix}\right|$ respectively.

A dual quaternion \cite{QiLingYan} can be defined as
$$
q=q_s+q_i\epsilon,
$$
where $q_s, q_i \in \mathbb{Q}$ are the standard part and the infinitesimal part of $q$, respectively. If $q_s\neq 0$, $q$ is appreciable, then the inverse of $q$ is $q^{-1}=q^{-1}_s-(q^{-1}_sq_iq^{-1}_s)\epsilon$, otherwise, $q$ is infinitesimal. For a dual quaternion $q=q_s+q_i\epsilon$, the conjugate of $q$ is $\overline{q}=\overline{q}_s+\overline{q}_i\epsilon$, and the magnitude \cite{QiLingYan} of $q$ is
$$
|q|=\left\{\begin{aligned}|q_s|+\displaystyle\frac{q_s\bar{q}_i+q_i\bar{q}_s}{2|q_s|}\epsilon, \  q_s\neq 0, \\
   |q_i|\epsilon,  \ \mathrm{otherwise}.\end{aligned}\right.
$$
The multiplication \cite{QiLingYan} of two dual quaternions $q=q_s+q_i\epsilon$, $p=p_s+p_i\epsilon$ are defined as
$$
qp=q_sp_s+(q_sp_i+q_ip_s)\epsilon.
$$
It is easy to verify that multiplication of dual quaternions satisfies distributive law.
A dual quaternion vector is denoted by $\mathbf{p}=\mathbf{p}_s+\mathbf{p}_i\epsilon$, where $\mathbf{p}_s, \mathbf{p}_i \in \mathbb{Q}^n$. For a dual quaternion vector $\mathbf{p} = (p_{(1)}, \cdots , p_{(n)})^T \in \mathbb{DQ}^n$, its 2-norm \cite{QiLingYan} is defined as
$$
\|\mathbf{p}\|_2=\left\{\begin{aligned}\sqrt{\sum_{t=1}^n|p_{(t)}|^2},  \ \mathrm{if} \ p_s\neq 0, \\
   \sqrt{\sum_{t=1}^n|p_{(t),i}|^2}\epsilon,  \ \mathrm{if} \ p_s=0.\end{aligned}\right.
$$

A dual quaternion matrix \cite{LingQiYan} $A=(a_{ij}) \in \mathbb{DQ}^{m\times n}$ can be denoted as
$$
A=A_s+A_i\epsilon,
$$
where $A_s=A_{11}+A_{12}\i+A_{13}\j+A_{14}\k, A_i=A_{21}+A_{22}\i+A_{23}\j+A_{24}\k \in \mathbb{Q}^{m\times n}$ are the standard part and the infinitesimal part of $A$, respectively. The conjugate, transpose, conjugate transpose \cite{LingQiYan} and $F^R$ norm \cite{CuiQi} of $A$ can be expressed as $\overline{A}=\overline{A}_s+\overline{A}_i\epsilon$, $A^T=A_s^T+A_i^T\epsilon$, $A^H=A_s^H+A_i^H\epsilon$, $\|A\|_{F^R}=\sqrt{\|A_s\|_F^2+\|A_i\|_F^2}$. A dual quaternion square matrix $A$ is referred to Hermitian \cite{LingHeQi} if $A^H=A$; Unitary if $A^HA=AA^H=I$.

Next, we will introduce the dual quaternion matrix LU decomposition, QR decomposition and singular value decomposition that are used in the process of zero-watermarking generation.

{\lemma \hskip-0.1mm\label{p2.1} \cite{WangLiWei} For $A \in \mathbb{DQ}^{n\times n}$, $A$ has a dual quaternion LU decomposition if $|A_s(1:t,1:t)|_q \neq 0$ for $t=1,2,\cdots, n-1$. If the dual quaternion LU decomposition exists and $|A_s|_q \neq 0$, then the dual quaternion LU decomposition is unique.
}

{\lemma \hskip-0.1mm\label{p2.2}\cite{ZhangLiWang} For given $A \in  \mathbb{DQ}^{m\times n}$, there exist a unitary matrix $Q \in  \mathbb{DQ}^{m\times m}$ and an upper triangular matrix $R \in  \mathbb{DQ}^{m\times n}$, such that,
$$
A=QR.
$$
}


{\lemma \hskip-0.1mm\label{p2.3} \cite{QiLuo} Suppose that $A = A_s + A_i\epsilon \in \mathbb{DQ}^{m\times n}$. Then there
exist dual quaternion unitary matrix $U \in \mathbb{DQ}^{m\times m}$ and dual quaternion unitary matrix $V \in \mathbb{DQ}^{n\times n}$, such that
$$
A=U \begin{bmatrix}\Sigma_t & 0 \\ 0 & 0 \end{bmatrix} V^H,
$$
where $\Sigma_t = diag(\sigma_1, \ldots, \sigma_r, \ldots ,\sigma_t) \in \mathbb{D}^{t\times t}$ is diagonal matrix, $ r \leq t \leq min\{m,n\}$, $\sigma_1 \geq \sigma_2 \geq \ldots \geq \sigma_r$ are positive appreciable dual numbers, and $\sigma_{r+1} \geq \ldots \geq \sigma_t$ are positive infinitesimal dual numbers.
}

\subsubsection{Color image model based on dual quaternion matrix}

The color image model based on the dual quaternion matrix refers to placing the R, G, and B channel information of the original carrier image in the three imaginary parts of the standard part of the dual quaternion matrix, and placing the R, G, and B channel information of the watermark image in the three imaginary parts of the dual part of the dual quaternion matrix, thereby forming a dual quaternion matrix
$$
F=F_1^{(1)}\i+F_1^{(2)}\j+F_1^{(3)}\k+(F_2^{(1)}\i+F_2^{(2)}\j+F_2^{(3)}\k)\epsilon.
$$

\subsubsection{The zero-watermarking generation process based on dual quaternion matrix decomposition}

Xu et al. investigated the LU decomposition \cite{XuWeiYan} and QR decomposition \cite{XuWeiWei} algorithms of dual matrix, they transformed the matrix decomposition problem into the problem of the solution of the real generalized Sylvester matrix equation $AX-YB=C$, and the existing conditions \cite{Baksalary}
\bea\label{2.1}
(I_m-AA^{\dag})C(I_n-B^{\dag}B)=O_{m\times n}\eea
of the solution were used to get the solution
\bea\label{2.2}
\left\{\begin{aligned}X&=A^{\dag}C+A^{\dag}ZB+(I_k-A^{\dag}A)W,\\
                      Y&=-(I_m-AA^{\dag})CB^{\dag}+Z-(I_m-AA^{\dag})ZBB^{\dag},\end{aligned}\right.\eea
where $W \in \mathbb{R}^{k\times n}$, $Z \in \mathbb{R}^{m\times l}$ being arbitrary, $A^{\dag}$ and $B^{\dag}$ are the Moore-Penrose inverse of $A$ and $B$.

In \cite{XuWeiYan}, the dual matrix LU decomposition is studied, for a dual matrix $A=A_s+A_i \epsilon \in \mathbb{D}^{m\times n}$, they transform the LU decomposition into two matrix equations
$$
\left\{\begin{aligned}A_s&=L_sU_s,\\
                      A_i&=L_sU_i+L_iU_s,\end{aligned}\right.
$$
the first equation can be achieved by performing the real matrix LU decomposition on the standard part of $A$. The second equation can be regarded as a Sylvester equation with unknowns $U_i$ and $L_i$. By using (\ref{2.2}), its solution can be obtained as
$$
\left\{\begin{array}{l}
U_{i}=L_{s}^{\dagger} A_{i}-P U_{s}, \\
L_{i}=\left(I_{m}-L_{s} L_{s}^{\dagger}\right) A_{i} U_{s}^{\dagger}+L_{s} P,
\end{array}\right.
$$
where $P=L_i^{\dag}Z$. Next, by leveraging the characteristics of $L_s$, $L_i$, $U_s$ and $U_i$, the elements in $P$ are determined, thereby obtaining $L_i$ and $U_i$. For the case of dual quaternions, they mentioned that it is necessary to determine the existence conditions for the quaternion Sylvester equation solution. In reference \cite{WangQW}, it was stated that these conditions are valid. Therefore, we can obtain the LU decomposition algorithm for the dual quaternion matrix in the same way. Then the dual part information $L_i$ and $U_i$ are used as the zero-watermarking information.

In \cite{XuWeiWei}, the dual matrix QR decomposition of dual matrix $A=A_s+A_i \epsilon \in \mathbb{D}^{m\times n}$ is studied, they obtained the dual component information
$$
\left\{\begin{aligned}Q_i&=Q_sP,\\
                      R_i&=Q_s^TA_i-PR_s,\end{aligned}\right.
$$
using a similar method, where $P=Q_s^TZ$ is an antisymmetric matrix. Unlike the dual matrix, when $A$ is a dual quaternion matrix, $P$ is no longer an antisymmetric matrix. Therefore, we will now discuss the situation of dual quaternion matrices.

For dual quaternion matrix $A=A_s+A_i \epsilon =(Q_s+Q_i\epsilon)(R_s+R_i\epsilon)\in \mathbb{DQ}^{m\times n}$, we can get
$$
\left\{\begin{aligned}A_s&=Q_sR_s,\\
                      A_i&=Q_sR_i+Q_iR_s,\end{aligned}\right.
$$
using (\ref{2.2}), we can obtain
$$
\left\{\begin{aligned}R_i&=Q_s^{\dag}A_i+Q_s^{\dag}Z(-R_s)+(I_m-Q_s^{\dag}Q_s)W,\\
                      Q_i&=-(I_m-Q_sQ_s^{\dag})A_i(-R_s)^{\dag}+Z -(I_m-Q_sQ_s^{\dag}Z(-R_s)(-R_s)^{\dag}).\end{aligned}\right.
$$
Since $Q$ is a unitary matrix and $Q^{\dag}=Q^H$ \cite{WeiLiZhang}, we can get
$$
\left\{\begin{aligned}R_i&=Q_s^HA_i-Q_s^HZR_s,\\ Q_i&=Z,\end{aligned}\right.
$$
let $P=Q_s^HZ=Q_s^HQ_i$, due to
$Q^HQ=(Q_s^H+Q_i^H\epsilon)(Q_s+Q_i\epsilon)=Q_s^HQ_s+(Q_s^HQ_i+Q_i^HQ_s)\epsilon=I_m$, then we can get $P+P^H=O_{m \times m}$, that is $P$ is anti-Hermitian matrix, which diagonal elements are pure imaginary quaternions. Let $B=Q_s^HA_i$, then $R_i=B-PR_s$, let $$P=\begin{bmatrix}\mathbf{p}_1 & \mathbf{p}_2 & \cdots & \mathbf{p}_n \end{bmatrix}, B=\begin{bmatrix}\mathbf{b}_1 & \mathbf{b}_2 & \cdots & \mathbf{b}_n \end{bmatrix},$$ $$R_i=\begin{bmatrix}\mathbf{r}_{i1} & \mathbf{r}_{i2} & \cdots & \mathbf{r}_{in} \end{bmatrix}=\begin{bmatrix}r_{i_{11}} & r_{i_{12}} & \cdots & r_{i_{1n}} \\  0 & r_{i_{22}} & \cdots & r_{i_{2n}} \\ \vdots & \vdots & \ddots & \vdots \\ 0  & 0 & \cdots & r_{i_{mn}} \end{bmatrix}, R_s=\begin{bmatrix}\mathbf{r}_{s1} & \mathbf{r}_{s2} & \cdots & \mathbf{r}_{sn} \end{bmatrix}=\begin{bmatrix}r_{s_{11}} & r_{s_{12}} & \cdots & r_{s_{1n}} \\  0 & r_{s_{22}} & \cdots & r_{s_{2n}} \\ \vdots & \vdots & \ddots & \vdots \\ 0  & 0 & \cdots & r_{s_{mn}} \end{bmatrix},$$
then
$$
\begin{bmatrix}\mathbf{p}_1 & \mathbf{p}_2 & \cdots & \mathbf{p}_n \end{bmatrix}
\begin{bmatrix}r_{s_{11}} & r_{s_{12}} & \cdots & r_{s_{1n}} \\  0 & r_{s_{22}} & \cdots & r_{s_{2n}} \\ \vdots & \vdots & \ddots & \vdots \\ 0  & 0 & \cdots & r_{s_{mn}} \end{bmatrix}
=\begin{bmatrix}\mathbf{b}_1 & \mathbf{b}_2 & \cdots & \mathbf{b}_n \end{bmatrix}
-\begin{bmatrix}r_{i_{11}} & r_{i_{12}} & \cdots & r_{i_{1n}} \\  0 & r_{i_{22}} & \cdots & r_{i_{2n}} \\ \vdots & \vdots & \ddots & \vdots \\ 0  & 0 & \cdots & r_{i_{mn}} \end{bmatrix}
$$
that is
$$
\left\{\begin{array}{ll}
\mathbf{p}_{1}r_{s_{11}} &=\mathbf{b}_{1}-\mathbf{r}_{i 1},\\
\mathbf{p}_{1}r_{s_{12}} + \mathbf{p}_{2}r_{s_{22}}&=\mathbf{b}_{2}-\mathbf{r}_{i 2},\\
&\vdots \\
\mathbf{p}_{1}r_{s_{1 k}} + \mathbf{p}_{2}r_{s_{2 k}} +\cdots+ \mathbf{p}_{k}r_{s_{k k}}&=\mathbf{b}_{k}-\mathbf{r}_{i k}, \\
&\vdots
\end{array}\right.
$$
Since $P$ is an anti-Hermitian matrix, we only need to determine its lower triangular part, that is
$$
\left\{\begin{array}{ll}
\mathbf{p}_{1}(1: m) & = \left[\mathbf{b}_{1}(1: m)-\begin{bmatrix}r_{i_{11}} & 0 & \cdots & 0 \end{bmatrix}^T \right]\frac{1}{r_{s_{11}}}, \\
\\
\mathbf{p}_{2}(2: m) & =\left[\mathbf{b}_{2}(2: m)-\begin{bmatrix}0 &r_{i_{22}} &  \cdots & 0 \end{bmatrix}^T-r_{s_{12}} \mathbf{p}_{1}(2: m)\right]\frac{1}{r_{s_{22}}}, \\
& \vdots \\
\mathbf{p}_{k}\left(k: m\right) & =\left[\mathbf{b}_{k}\left(k: m\right)-\begin{bmatrix}0 \cdots & 0 &r_{i_{kk}} & 0 &  \cdots & 0 \end{bmatrix}^T-\sum_{t=1}^{k-1} r_{s_{t k}} \mathbf{p}_{t}\left(k: m\right)\right]\frac{1}{r_{s_{kk}}},\\
& \vdots
\end{array}\right.
$$
The diagonal elements of $P$ are pure imaginary quaternions, while the diagonal elements of $R_s$ and $R_i$ are real numbers. Therefore, when determining the diagonal elements of $P$, we only need to take the imaginary part of the diagonal elements of the right-hand term. Once $P$ is determined, we can obtain the dual part information $Q_i$ and $R_i$. Then $Q_i$ and $R_i$ can be used as the zero-watermarking information.


Qi et al. \cite{QiLuo} proposed the dual quaternion matrix singular value decomposition algorithm, which is based on the eigenvalue decomposition of the dual quaternion matrix. For dual quaternion matrix $A=A_s+A_i \epsilon \in \mathbb{DQ}^{m\times n}$, the following singular value decomposition holds, that is
$$
A=U \begin{bmatrix}\Sigma_t & 0 \\ 0 & 0 \end{bmatrix} V^H,
$$
where
$U=\hat{U}\begin{bmatrix}I_r &    \\  & W_2\end{bmatrix}
=\begin{bmatrix}\hat{U}_1 & \hat{U}_2\end{bmatrix} \begin{bmatrix}I_r &   \\ & W_2\end{bmatrix}
=\begin{bmatrix}\hat{U}_1 & \hat{U}_2 W_2\end{bmatrix}$, $V=\hat{V}\begin{bmatrix}I_r &  \\  & W_1\end{bmatrix}
=\begin{bmatrix}\hat{V}_1 & \hat{V}_2\end{bmatrix} \begin{bmatrix}I_r &    \\  & W_1\end{bmatrix}
=\begin{bmatrix}\hat{V}_1 & \hat{V}_2 W_1 \end{bmatrix}$,
$\hat{U}_{1}=\hat{U}_{:, 1: r}$, $\hat{U}_{2}=\hat{U}_{:, r+1: n}$, $\hat{V}_{1}=A \hat{U}_{1} \Sigma_{r}^{-1}$, $\hat{V}_{2}$ and $ \hat{V}_{1}$ form an unitary matrix, and $\hat{V}^H A \hat{U}=\left[\begin{array}{cc}\Sigma_{r} & 0 \\ 0 & G \epsilon\end{array}\right]$, $W_{1}^H G W_{2}=D \in \mathbb{Q}^{(m-r) \times(n-r)}$, $D_{ij}=0$, $i\neq j$, $D_{ii}\geqslant 0$, $i=1, \cdots, \min \{m-r, n-r\}$.
Use the dual part information $U_i$, $\Sigma_i$ and $V_i$ as the zero-watermarking information, and $W_1$, $W_2$ as the keys.



\subsubsection{The watermark verification process}

During the watermark verification process, we first reconstruct the watermark image, and then calculate various indicators between the reconstructed watermark image and the original watermark image. When generating the zero-watermarking, we place the original carrier image features in the standard part of the dual quaternion matrix and place the watermark image in the dual part. Therefore, the watermark reconstruction is to reconstruct the dual part of the dual quaternion matrix using the image to be verified and the zero-watermarking. Next, we will introduce the watermark image reconstruction process of the fragile zero-watermarking method based on three different matrix decomposition respectively.

For the fragile zero-watermarking method based on DQLU decomposition, since $A=A_s+A_i \epsilon=(L_s+L_i \epsilon)(U_s+U_i \epsilon)=L_sU_s+(L_sU_i+L_iU_s)\epsilon$, to obtain $A_i$, we need to obtain $L_s$ and $U_s$. Represent the image to be verified as a quaternion matrix by utilizing the quaternion color image model \cite{PeiCheng}, and perform the quaternion matrix LU (QLU) decomposition on it, $A_s = L_sU_s$, then use zero-watermarking $L_i$ and $U_i$, we can get $A_i=L_sU_i+L_iU_s$.

For the fragile zero-watermarking method based on DQQR decomposition, similar to the DQLU decomposition, the standard part is first subjected to quaternion QR (QQR) decomposition to obtain $Q_s$ and $R_s$, and then the zero-watermarking information $Q_i$ and $R_i$ are utilized to obtain $A_i=Q_sR_i+Q_iR_s$.

For the fragile zero-watermarking method based on DQSVD decomposition, since $A=A_s+A_i \epsilon=(U_s+U_i \epsilon)(\Sigma_s+\Sigma_i \epsilon)(V_s^H+V_i^H \epsilon)=U_s\Sigma_sV_s^H+(U_s\Sigma_sV_i^H+U_s\Sigma_iV_s^H+U_i\Sigma_sV_s^H)\epsilon$, to obtain $A_i$, we need to obtain $U_s$, $\Sigma_s$ and $V_s$. Represent the image to be verified as a quaternion matrix by utilizing the quaternion color image model, and perform the quaternion matrix SVD (QSVD) decomposition on it, $A_s = \hat{U}_s\hat{\Sigma}_s\hat{V}_s^H$, combine the obtained $\hat{U}_s$ and $\hat{V}_s$ with the keys $W_1$ and $W_2$ to obtain $U_s$, $\Sigma_s$ and $V_s$, then use zero-watermarking $U_i$, $\Sigma_i$ and $V_i$, we can get $A_i=U_s\Sigma_sV_i^H+U_s\Sigma_iV_s^H+U_i\Sigma_sV_s^H$.

During the verification stage, we determine whether the original carrier image has been attacked by calculating the signal-to-noise ratio (PSNR), structural similarity index (SSIM), the bit error rate (BER) and normalized correlation coefficient (NC) between the restored watermark image and the original watermark image.

The PSNR can be defined as
$$
PSNR=10 lg \displaystyle\frac{3N^2(maxF(x,y,k))^2}{\sum\limits_{x=1}^N \sum\limits_{y=1}^N \sum\limits_{k=1}^3(F(x,y,k)-\widetilde{F}(x,y,k))^2},
$$
where $max F(x, y, k)$ is 255 (the maximum pixel value for color image). $F(x, y, k)$ and $\widetilde{F}(x, y, k)$ are pixel values in original watermark image $\mathcal{F}$ and the restored watermark image $\tilde{\mathcal{F}}$ at $(x, y, k)$ position, respectively.

The SSIM can be defined as
$$
SSIM(F,\widetilde{F})=\displaystyle\frac{(2\mu_F \mu_{\widetilde{F}}+c_1)(2\sigma_{F\widetilde{F}}+c_2)}
{(\mu_F^2 \mu_ {\widetilde{F}}^2+c_1)(\sigma_F^2 \sigma_{\widetilde{F}}^2+c_2)},
$$
where $F$ and $\widetilde{F}$ denote the original watermark image $\mathcal{F}$ and the restored watermark image $\tilde{\mathcal{F}}$ respectively, $\mu_F$ and $\mu_ {\widetilde{F}}$ are the average values of $F$ and $\widetilde{F}$ respectively, $\sigma_{F\widetilde{F}}$ is the covariance of $F$ and $\widetilde{F}$, $\sigma_F$ and $\sigma_{\widetilde{F}}$ are the variances of $F$ and $\widetilde{F}$ respectively, $c_1$ and $c_2$ are the two variables to stabilize the division with the weak denominator.

The BER can be defined as
$$
BER=\displaystyle\frac{f}{3 m n},
$$
where $f$ is the incorrect number of extracted bits, and $mn$ represents the length of the tested watermarking bits.

The NC can be defined as
$$
NC=\displaystyle\frac{\sum\limits_{x=1}^N \sum\limits_{y=1}^N \sum\limits_{k=1}^3 G(x,y,k) \widetilde{G}(x,y,k)}{\sqrt{\sum\limits_{x=1}^N \sum\limits_{y=1}^N \sum\limits_{k=1}^3 G(x,y,k)^2} \sqrt{\sum\limits_{x=1}^N \sum\limits_{y=1}^N \sum\limits_{k=1}^3 \widetilde{G}(x,y,k)^2}},
$$
where $N$ is the height (width) of the tested watermarking, $G(x, y, k)$ and $\widetilde{G}(x, y, k)$ are pixel values in original watermark image $\mathcal{G}$ and restored watermark image $\tilde{\mathcal{G}}$ at $(x, y, k)$ position, respectively.

\section{Fragile zero-watermarking method based on dual quaternion matrix decomposition}

\subsection{Fragile zero-watermarking method}

In the second section, the preprocessing process, zero-watermarking generation process and watermark verification process of fragile zero-watermarking methods based on DQLU decomposition, DQQR decomposition and DQSVD decomposition are analyzed respectively. Next, the zero-watermarking generation algorithm and the watermark verification algorithm are presented respectively. First, we talk about the zero-watermarking generation algorithm.

\begin{breakablealgorithm}
  \caption{The algorithm for zero-watermarking generation.}
  \label{code:recentEnd1}
  \begin{algorithmic}[1]
\item[$\mathbf{Step \;1}$] The original carrier image is preprocessed by using the fast Fourier transform, converting the image from the spatial domain to the frequency domain to obtain the feature image. The watermark image is encrypted using Arnold scrambling, resulting in an encrypted image, and the scrambling count is saved as key.
\item[$\mathbf{Step \;2}$] Place the feature image and the encrypted image in the standard part and the dual part of the dual quaternion matrix respectively, thereby forming the dual quaternion matrix.
\item[$\mathbf{Step \;3}$] Perform matrix decomposition on the obtained dual quaternion matrix, and use the obtained dual part information as the zero-watermarking.
\end{algorithmic}
\end{breakablealgorithm}

\begin{figure}[H]
\centering
\includegraphics[width=15cm]{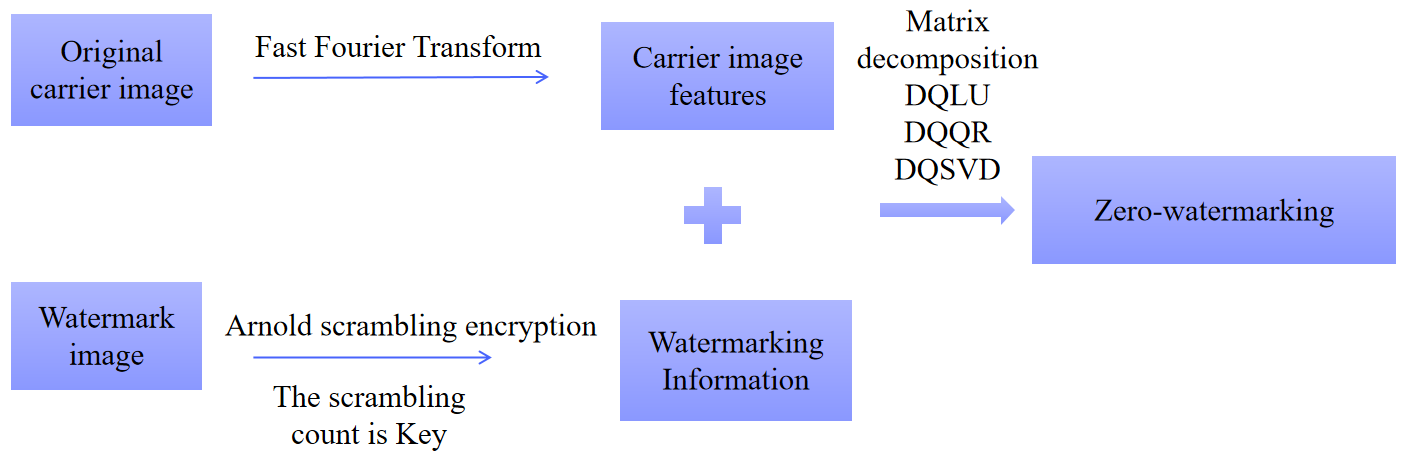}
\caption{Zero-watermarking generation process.}
\label{f3.1}
\end{figure}

A flowchart (Fig. \ref{f3.1}) can be used to describe zero-watermarking generation process more clearly.

Next we introduce the watermark verification algorithm.

\begin{breakablealgorithm}
  \caption{The algorithm for watermark verification.}
  \label{code:recentEnd2}
  \begin{algorithmic}[1]
\item[$\mathbf{Step \;1}$] The image to be verified is preprocessed by using the fast Fourier transform, converting the image from the spatial domain to the frequency domain to obtain the feature image.
\item[$\mathbf{Step \;2}$] Represent the feature image as a pure imaginary quaternion matrix.
\item[$\mathbf{Step \;3}$] Perform matrix decomposition on the obtained quaternion matrix, combine the obtained matrices with zero-watermarking to get the encrypted watermark image.
\item[$\mathbf{Step \;4}$] Using key and anti-Arnold scrambling to restore the encrypted image, then watermark image is obtained.
\end{algorithmic}
\end{breakablealgorithm}

\begin{figure}[H]
\centering
\includegraphics[width=15cm]{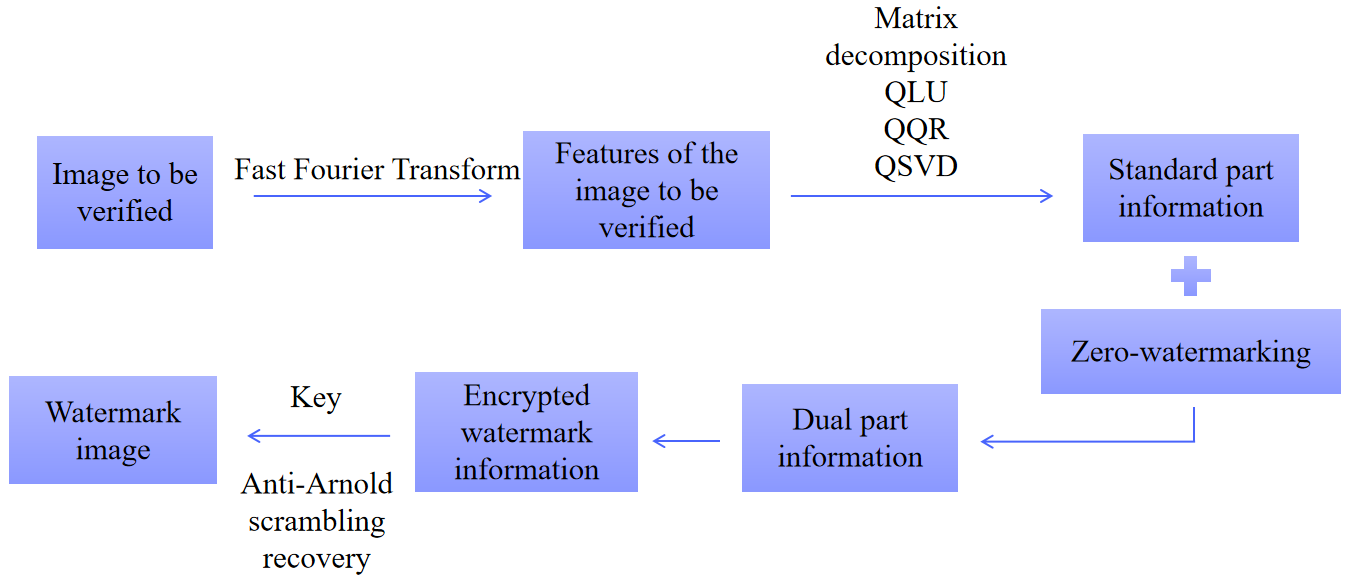}
\caption{Watermark verification process.}
\label{f3.2}
\end{figure}

The flowchart (Fig. \ref{f3.2}) for the watermark verification process can be described as follows.

\subsection{Numerical experiment}

\begin{figure}[H]
\centering
\subfigure[001.]
{
\includegraphics[width=2.5cm]{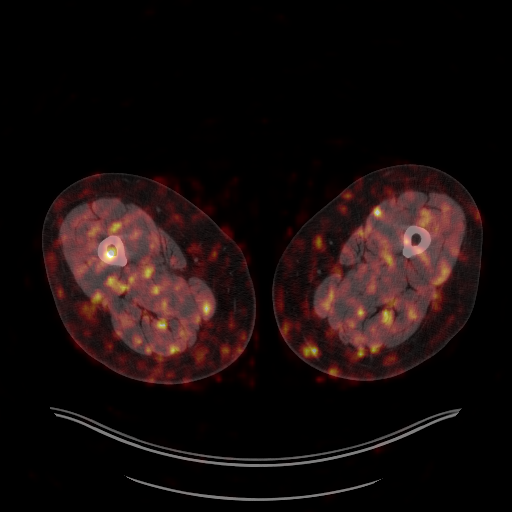}
}
\hskip .05cm
\subfigure[018.]
{
\includegraphics[width=2.5cm]{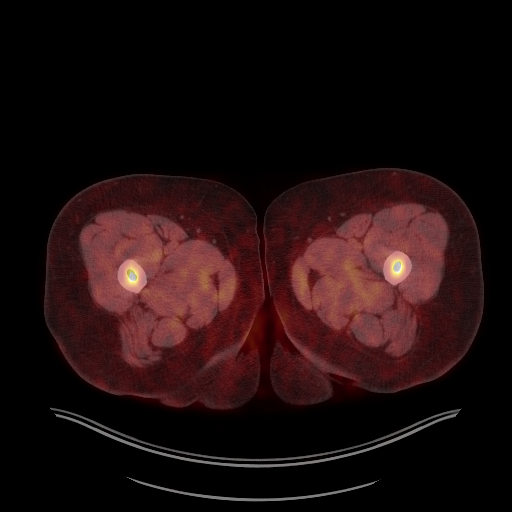}
}
\hskip .05cm
\subfigure[035.]
{
\includegraphics[width=2.5cm]{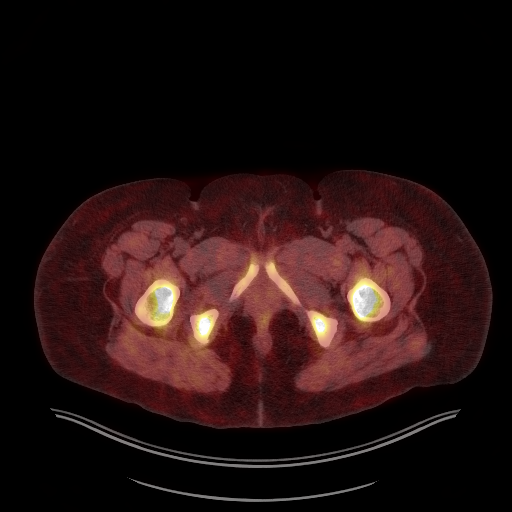}
}
\hskip .05cm
\subfigure[052.]
{
\includegraphics[width=2.5cm]{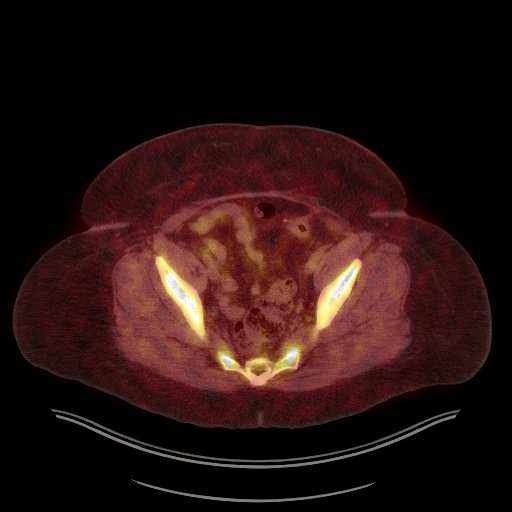}
}
\hskip .05cm
\subfigure[069.]
{
\includegraphics[width=2.5cm]{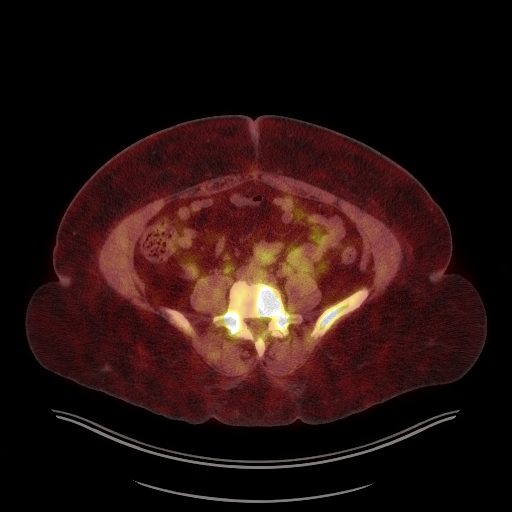}
}
\subfigure[086.]
{
\includegraphics[width=2.5cm]{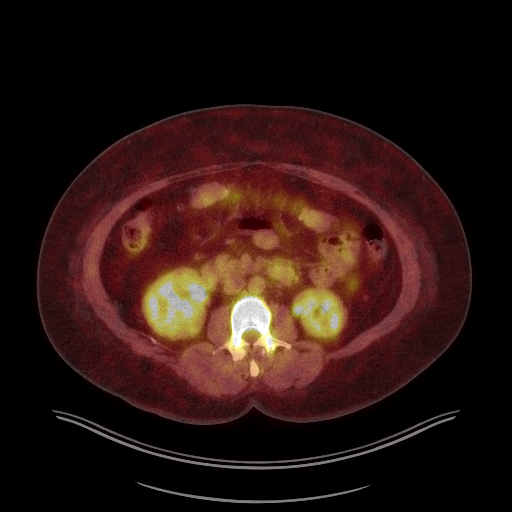}
}
\hskip .05cm
\subfigure[103.]
{
\includegraphics[width=2.5cm]{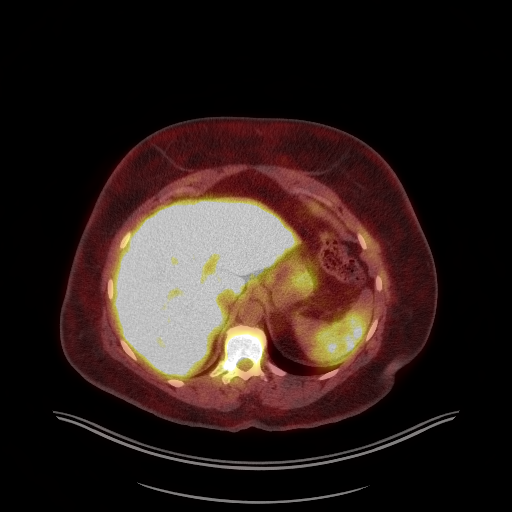}
}
\hskip .05cm
\subfigure[120.]
{
\includegraphics[width=2.5cm]{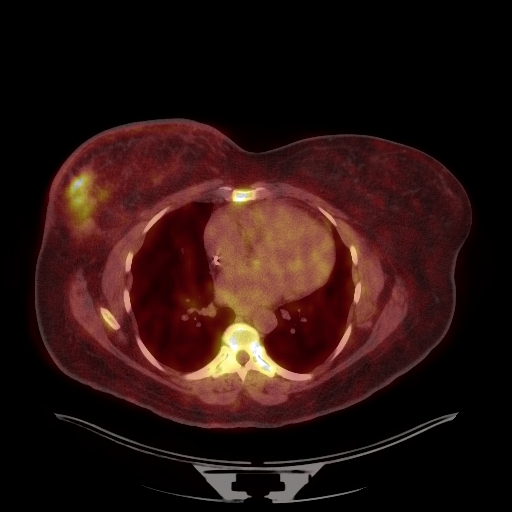}
}
\hskip .05cm
\subfigure[137.]
{
\includegraphics[width=2.5cm]{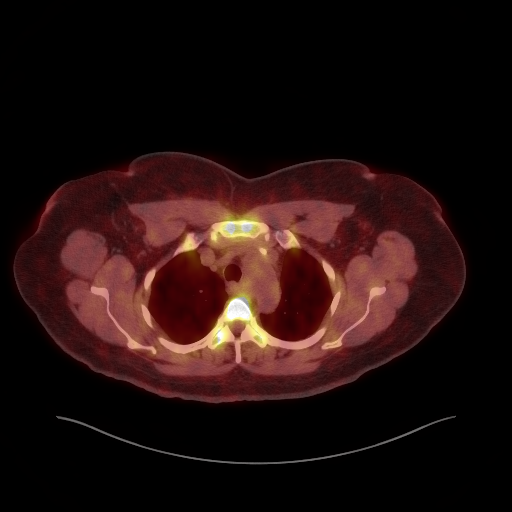}
}
\hskip .05cm
\subfigure[145.]
{
\includegraphics[width=2.5cm]{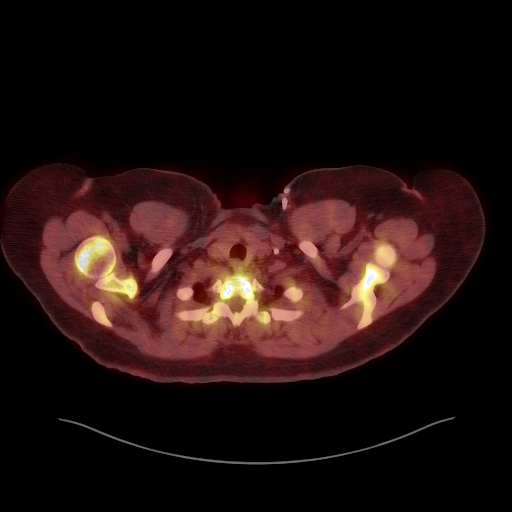}
}
\caption{$512\times 512$ medical images.}
\label{f3.3}
\end{figure}

First, we verify the feasibility of our method. We randomly select ten images from The Cancer Imaging Archive dataset \cite{Rutherford} (Fig. \ref{f3.3}) and generate zero-watermarking using the fragile zero-watermarking generation algorithms based on DQLU decomposition, DQQR decomposition and DQSVD decomposition respectively. Without applying any attacks, we extract and verify the watermark images using the corresponding fragile watermark verification algorithms based on QLU decomposition, QQR decomposition and QSVD decomposition. And calculate the indicators PSNR, SSIM, NC and BER, all of which can completely extract the watermark. Moreover, PSNR is Inf, SSIM and NC are 1, and BER is 0.

\begin{figure}[H]
\centering
\includegraphics[width=2.5cm]{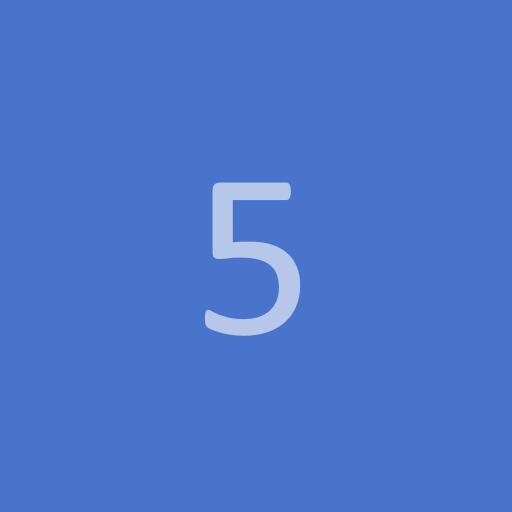}
\caption{Watermark image.}
\label{f3.4}
\end{figure}

Secondly, we randomly selected four groups of similar images from The Cancer Imaging Archive dataset, such as 002 and 003. It is difficult to detect the differences between them with the naked eye. We used the fragile zero-watermarking generation algorithm to generate a zero-watermarking for 002, and then took 003 as the image to be verified. We extracted and verified it with the watermark verification algorithm, and obtained Table \ref{tab1}. It can be seen from the table that the watermark images cannot be extracted, which proves the accuracy of our algorithm.

\begin{table}[H]
  \centering
  \caption{Experimental results of two similar medical images.}
	\label{tab1}
  \begin{tabular}{ccccccc}
    \hline\hline
    \centering
  Original image    & Image to be verified & DQLU  & DQQR  & DQSVD  \\
    \hline
     & & & & \\
    \centering \begin{minipage}[b]{0.15\columnwidth}
		\centering
		\raisebox{-.5\height}{\includegraphics[width=1.5cm,height=1.5cm]{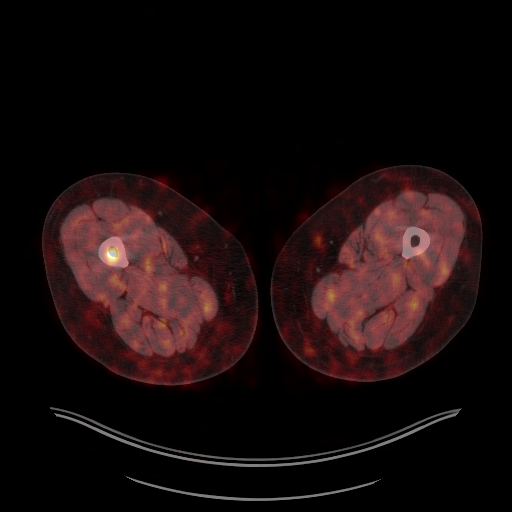}}
	\end{minipage}
& \begin{minipage}[b]{0.15\columnwidth}
		\centering
		\raisebox{-.5\height}{\includegraphics[width=1.5cm,height=1.5cm]{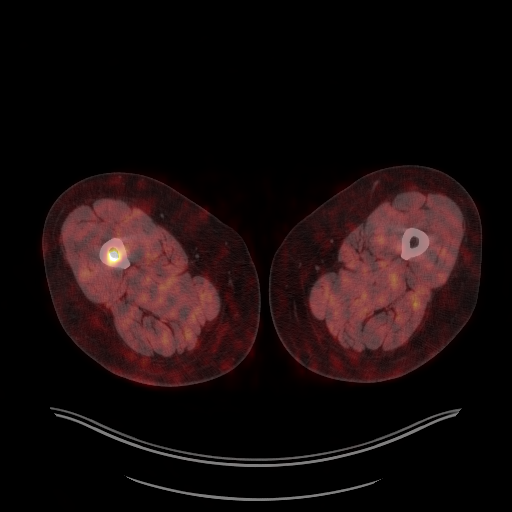}}
	\end{minipage}
    & \begin{minipage}[b]{0.15\columnwidth}
		\centering
		\raisebox{-.5\height}{\includegraphics[width=1.5cm,height=1.5cm]{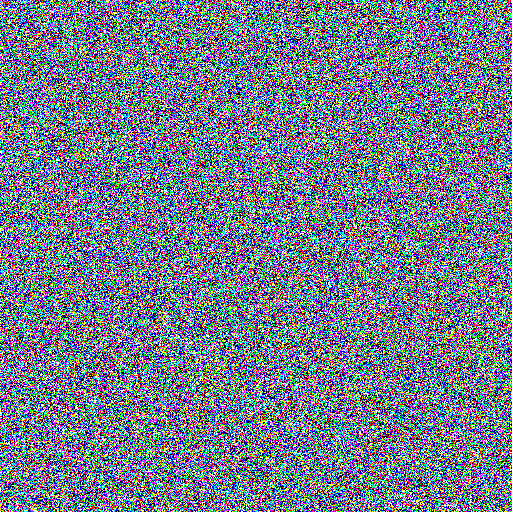}}
	\end{minipage}
    & \begin{minipage}[b]{0.15\columnwidth}
		\centering
		\raisebox{-.5\height}{\includegraphics[width=1.5cm,height=1.5cm]{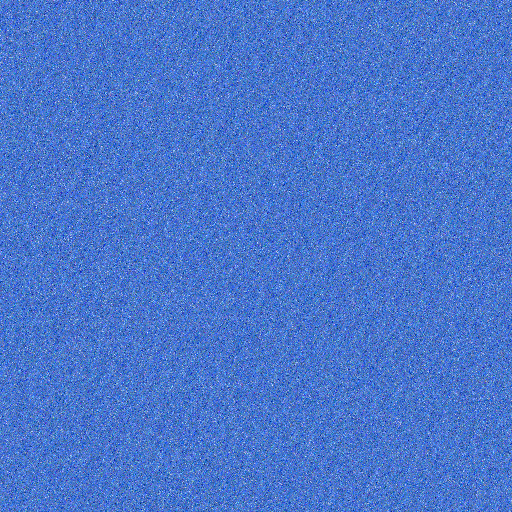}}
	\end{minipage}
    & \begin{minipage}[b]{0.15\columnwidth}
		\centering
		\raisebox{-.5\height}{\includegraphics[width=1.5cm,height=1.5cm]{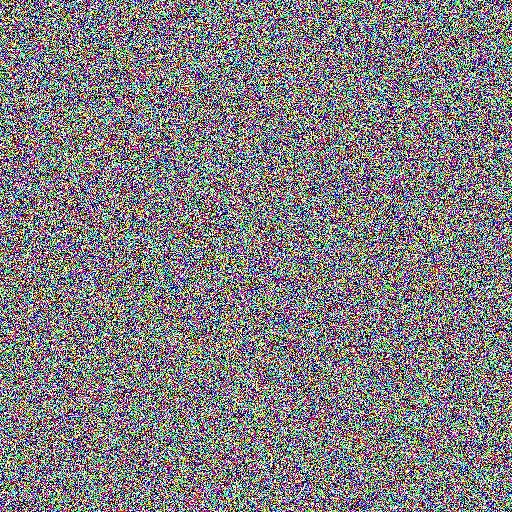}}
	\end{minipage}
    \\
    \centering 002& 003 &0.4877/0.7007&0.4231/0.9799&0.5009/0.6613 \\
    \centering    &     &6.1038/0.0708&18.8903/0.8765&5.3912/0.0045 \\
    \hline
     & & & & \\
    \centering \begin{minipage}[b]{0.15\columnwidth}
		\centering
		\raisebox{-.5\height}{\includegraphics[width=1.5cm,height=1.5cm]{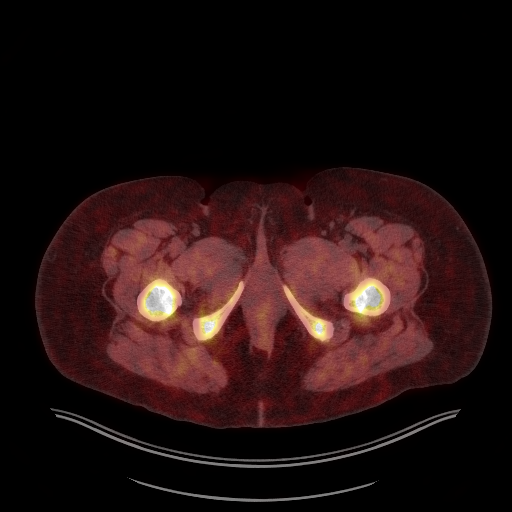}}
	\end{minipage}
& \begin{minipage}[b]{0.15\columnwidth}
		\centering
		\raisebox{-.5\height}{\includegraphics[width=1.5cm,height=1.5cm]{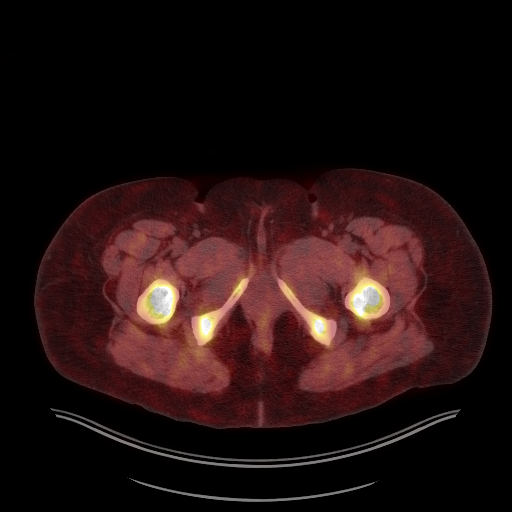}}
	\end{minipage}
    & \begin{minipage}[b]{0.15\columnwidth}
		\centering
		\raisebox{-.5\height}{\includegraphics[width=1.5cm,height=1.5cm]{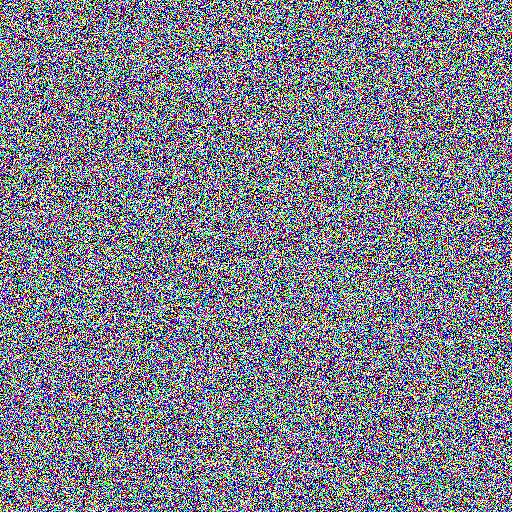}}
	\end{minipage}
    & \begin{minipage}[b]{0.15\columnwidth}
		\centering
		\raisebox{-.5\height}{\includegraphics[width=1.5cm,height=1.5cm]{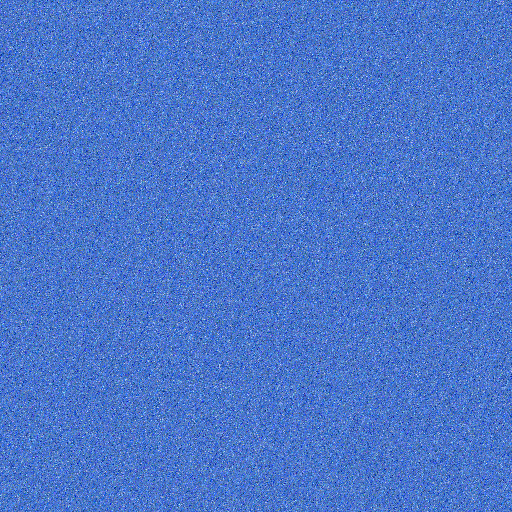}}
	\end{minipage}
    & \begin{minipage}[b]{0.15\columnwidth}
		\centering
		\raisebox{-.5\height}{\includegraphics[width=1.5cm,height=1.5cm]{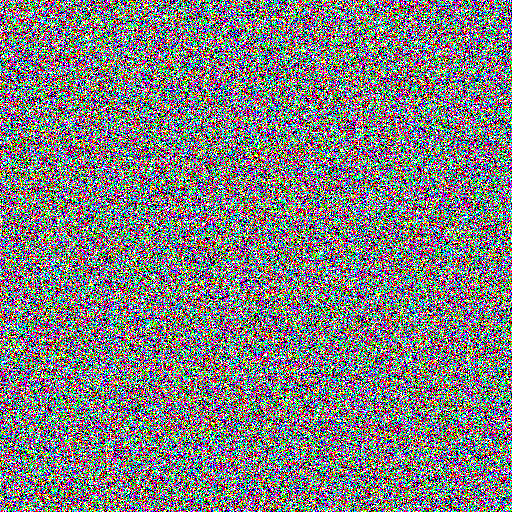}}
	\end{minipage}
    \\
    \centering 033& 034 &0.4947/0.6996&0.4233/0.9793&0.4998/0.6530 \\
    \centering    &     &5.9570/0.0422&18.8022/0.8717&5.2804/0.0047 \\
    \hline
    & & & & \\
    \centering \begin{minipage}[b]{0.15\columnwidth}
		\centering
		\raisebox{-.5\height}{\includegraphics[width=1.5cm,height=1.5cm]{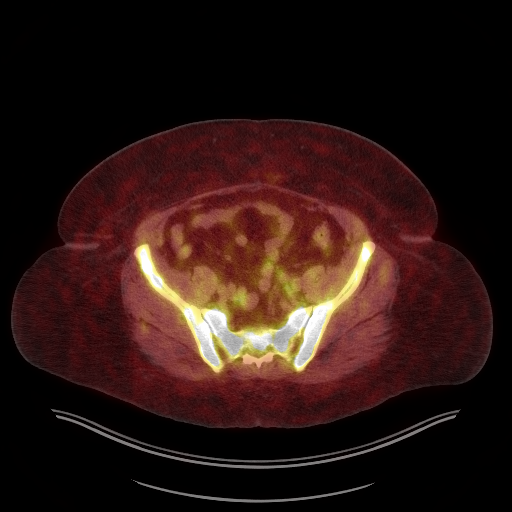}}
	\end{minipage}
& \begin{minipage}[b]{0.15\columnwidth}
		\centering
		\raisebox{-.5\height}{\includegraphics[width=1.5cm,height=1.5cm]{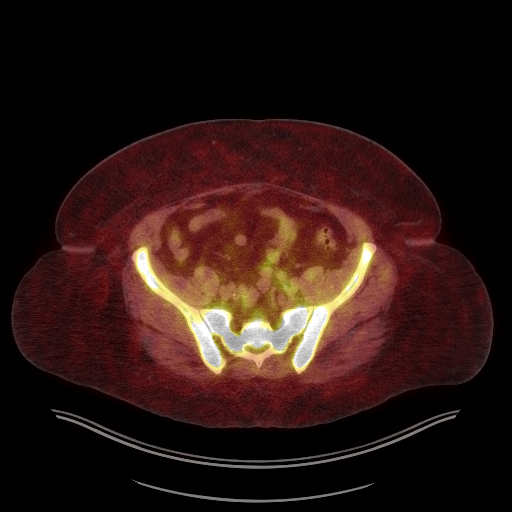}}
	\end{minipage}
    & \begin{minipage}[b]{0.15\columnwidth}
		\centering
		\raisebox{-.5\height}{\includegraphics[width=1.5cm,height=1.5cm]{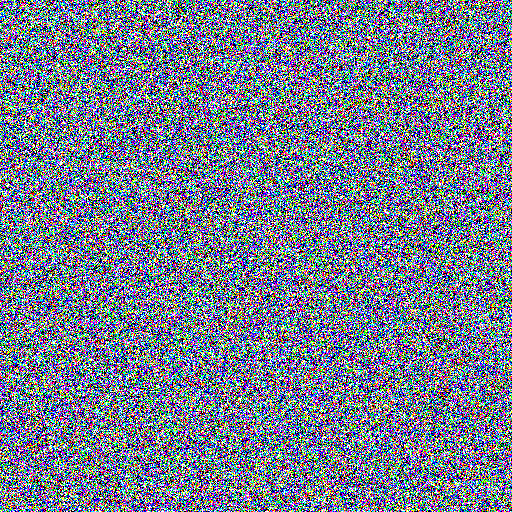}}
	\end{minipage}
    & \begin{minipage}[b]{0.15\columnwidth}
		\centering
		\raisebox{-.5\height}{\includegraphics[width=1.5cm,height=1.5cm]{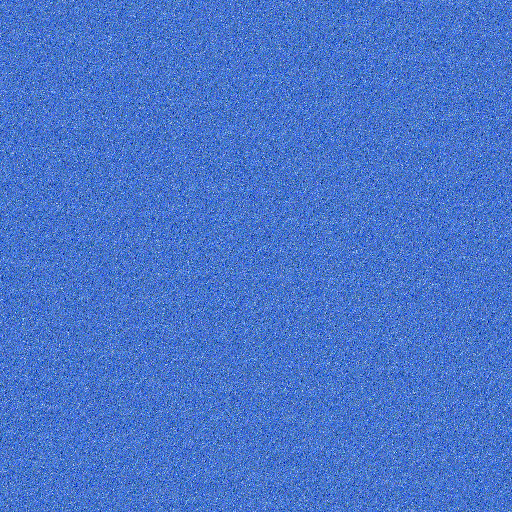}}
	\end{minipage}
    & \begin{minipage}[b]{0.15\columnwidth}
		\centering
		\raisebox{-.5\height}{\includegraphics[width=1.5cm,height=1.5cm]{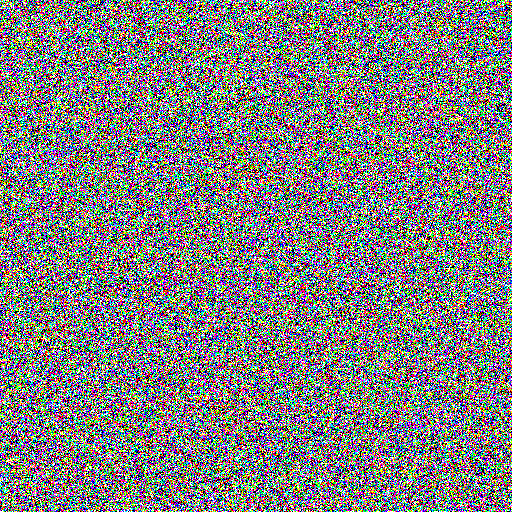}}
	\end{minipage}
    \\
    \centering 058& 059 &0.4888/0.7010&0.4269/0.9781&0.5003/0.6524 \\
    \centering    &     &6.1022/0.0639&18.5514/0.8663&5.2713/0.0022 \\
    \hline
     & & & & \\
    \centering \begin{minipage}[b]{0.15\columnwidth}
		\centering
		\raisebox{-.5\height}{\includegraphics[width=1.5cm,height=1.5cm]{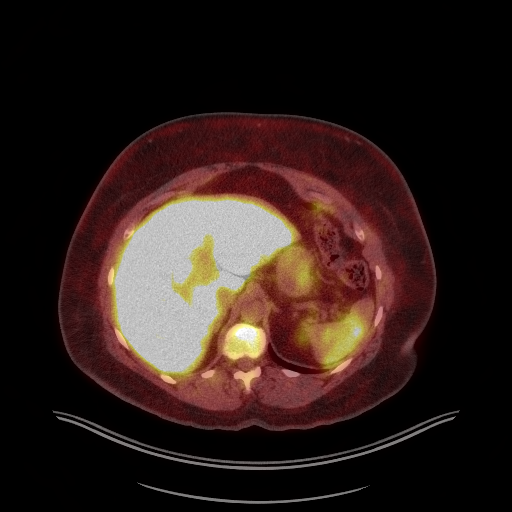}}
	\end{minipage}
& \begin{minipage}[b]{0.15\columnwidth}
		\centering
		\raisebox{-.5\height}{\includegraphics[width=1.5cm,height=1.5cm]{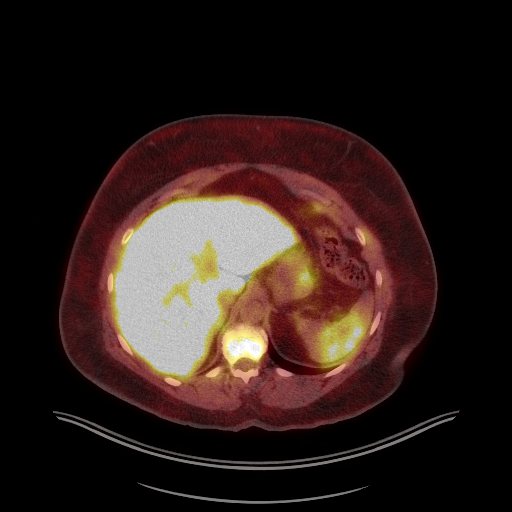}}
	\end{minipage}
    & \begin{minipage}[b]{0.15\columnwidth}
		\centering
		\raisebox{-.5\height}{\includegraphics[width=1.5cm,height=1.5cm]{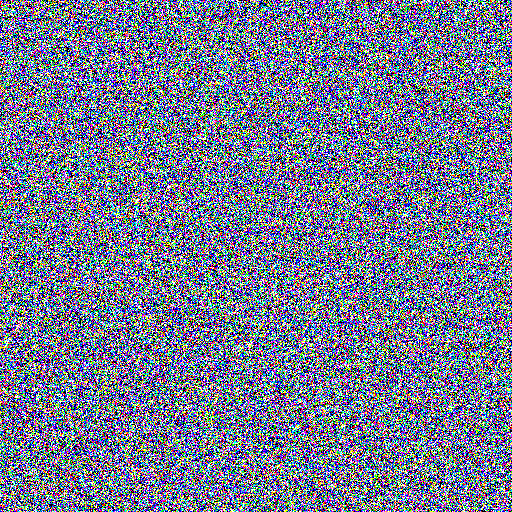}}
	\end{minipage}
    & \begin{minipage}[b]{0.15\columnwidth}
		\centering
		\raisebox{-.5\height}{\includegraphics[width=1.5cm,height=1.5cm]{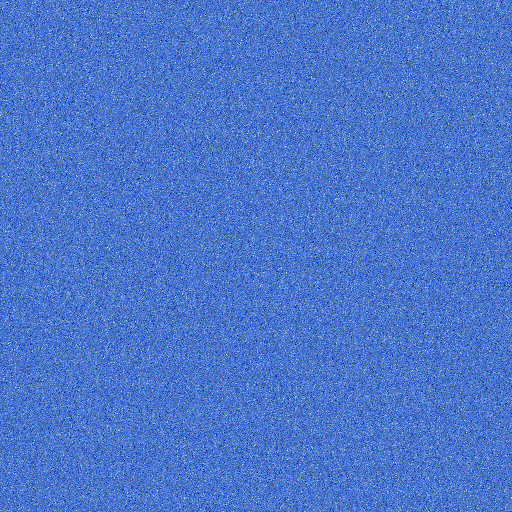}}
	\end{minipage}
    & \begin{minipage}[b]{0.15\columnwidth}
		\centering
		\raisebox{-.5\height}{\includegraphics[width=1.5cm,height=1.5cm]{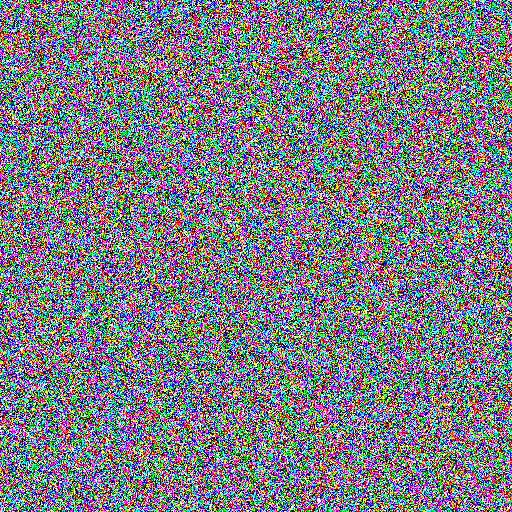}}
	\end{minipage}
    \\
    \centering 101& 102 &0.4879/0.7027&0.4229/0.9796&0.4953/0.6713 \\
    \centering    &     &6.1291/0.0715&18.8340/0.8730&5.6206/0.0282 \\
    \hline
  \end{tabular}
\end{table}

Finally, we conducted an anti-attack experiment. Taking 001 and 061 in The Cancer Imaging Archive dataset as examples, we respectively applied Gaussian noise (0, 0.01), $90\%$ compression attack, Counterclockwise rotation $3^o$, Center cropping attack, Brightening attack, Modify a single pixel to the images to be verified, and calculated their various indicators to obtain Table \ref{tab2}-\ref{tab3}. It can be seen that after the attack, the performance of each indicator of the watermark image was very poor. Therefore, our algorithm can be used to verify whether medical images have been attacked.

All these computation are performed on an Intel(R) Core(TM) i9-13900HX @2.20GHz/32GB computer.

\section{Conclusion}

We propose a fragile zero-watermarking method based on dual quaternion matrices decomposition for copyright protection and content tampering detection. This method encrypts the patient information image through Arnold transformation, extracts the features of medical images using the fast Fourier transform, correlates the medical images with the patient information images by operating on the standard part and dual part of the dual quaternions, and generates zero-watermark information based on the characteristics of the dual quaternion matrix DQLU, DQQR and DQSVD decomposition. Ultimately, numerical experiments prove that this method can be used for  copyright protection and content tampering detection of medical images.

\begin{table}[H]
\centering
\caption{ Experimental results under different attack with medical image 001.}
	\label{tab2}
	\begin{tabular}{ccccc}
 \hline\hline
 \multicolumn{5}{c}{Fragile zero-watermarking method based on DQLU decomposition} \\
 \hline
  Attack                           & BER    & NC     & PSNR   & SSIM \\
 \hline
 Gaussian noise (0, 0.01)          & 0.4906 & 0.7160 & 6.0294 & 0.0672 \\
 $90\%$ compression attack         & 0.4912 & 0.6936 & 5.9582 & 0.0489 \\
 Counterclockwise rotation $3^o$   & 0.4911 & 0.6986 & 5.9694 & 0.0522 \\
 Center cropping attack            & 0.4870 & 0.6905 & 5.9693 & 0.0533 \\
 Brightening attack                & 0.4883 & 0.6976 & 6.0019 & 0.0571 \\
 Modify a single pixel             & 0.4128 & 0.7863 & 7.8858 & 0.2003 \\
 \hline
 \hline
 \multicolumn{5}{c}{Fragile zero-watermarking method based on DQQR decomposition} \\
 \hline
  Attack                           & BER    & NC     & PSNR   & SSIM \\
 \hline
 Gaussian noise (0, 0.01)          & 0.4381 & 0.9728 & 17.0588 & 0.8571 \\
 $90\%$ compression attack         & 0.4199 & 0.9808 & 19.1288 & 0.8802 \\
 Counterclockwise rotation $3^o$   & 0.4218 & 0.9798 & 18.9007 & 0.8760 \\
 Center cropping attack            & 0.3843 & 0.9851 & 20.2401 & 0.9028 \\
 Brightening attack                & 0.4151 & 0.9820 & 19.3649 & 0.8891 \\
 Modify a single pixel             & 0.3571 & 0.9869 & 20.7905 & 0.9119 \\
 \hline
 \hline
 \multicolumn{5}{c}{Fragile zero-watermarking method based on DQSVD decomposition} \\
 \hline
  Attack                           & BER    & NC     & PSNR   & SSIM \\
 \hline
 Gaussian noise (0, 0.01)          & 0.5001 & 0.6532 & 5.3301 & 0.0038 \\
 $90\%$ compression attack         & 0.5000 & 0.6574 & 5.3320 & 0.0044 \\
 Counterclockwise rotation $3^o$   & 0.5001 & 0.6573 & 5.3296 & 0.0044 \\
 Center cropping attack            & 0.4986 & 0.6582 & 5.4131 & 0.0107 \\
 Brightening attack                & 0.5002 & 0.6572 & 5.3297 & 0.0038 \\
 Modify a single pixel             & 0.5009 & 0.6540 & 5.3105 & 0.0008 \\
 \hline
\end{tabular}
\end{table}

\begin{table}[H]
\centering
\caption{ Experimental results under different attack with medical image 061.}
	\label{tab3}
	\begin{tabular}{ccccc}
 \hline\hline
 \multicolumn{5}{c}{Fragile zero-watermarking method based on DQLU decomposition} \\
 \hline
  Attack                           & BER    & NC     & PSNR   & SSIM \\
 \hline
 Gaussian noise (0, 0.01)          & 0.4924 & 0.7188 & 6.0631 & 0.0646 \\
 $90\%$ compression attack         & 0.4841 & 0.6948 & 6.0865 & 0.0760 \\
 Counterclockwise rotation $3^o$   & 0.4936 & 0.7021 & 5.9793 & 0.0442 \\
 Center cropping attack            & 0.4844 & 0.6991 & 6.1051 & 0.0648 \\
 Brightening attack                & 0.4828 & 0.7002 & 6.1521 & 0.0857 \\
 Modify a single pixel             & 0.4195 & 0.7412 & 6.8925 & 0.1216 \\
 \hline
 \hline
 \multicolumn{5}{c}{Fragile zero-watermarking method based on DQQR decomposition} \\
 \hline
  Attack                           & BER    & NC     & PSNR   & SSIM \\
 \hline
 Gaussian noise (0, 0.01)          & 0.4340 & 0.9722 & 17.9455 & 0.8758 \\
 $90\%$ compression attack         & 0.4169 & 0.9818 & 19.3555 & 0.8869 \\
 Counterclockwise rotation $3^o$   & 0.4226 & 0.9799 & 18.9236 & 0.8768 \\
 Center cropping attack            & 0.4087 & 0.9829 & 19.6294 & 0.8940 \\
 Brightening attack                & 0.4101 & 0.9831 & 19.6709 & 0.8958 \\
 Modify a single pixel             & 0.3487 & 0.9863 & 20.6008 & 0.9088 \\
 \hline
 \hline
 \multicolumn{5}{c}{Fragile zero-watermarking method based on DQSVD decomposition} \\
 \hline
  Attack                           & BER    & NC     & PSNR   & SSIM \\
 \hline
 Gaussian noise (0, 0.01)          & 0.4999 & 0.6518 & 5.2900 & 0.0028 \\
 $90\%$ compression attack         & 0.4997 & 0.6541 & 5.3013 & 0.0058 \\
 Counterclockwise rotation $3^o$   & 0.5000 & 0.6540 & 5.2936 & 0.0039 \\
 Center cropping attack            & 0.4996 & 0.6545 & 5.3033 & 0.0059 \\
 Brightening attack                & 0.4999 & 0.6541 & 5.2967 & 0.0046 \\
 Modify a single pixel             & 0.5002 & 0.6541 & 5.2928 & 0.0039 \\
 \hline
\end{tabular}
\end{table}

\section*{Conflict of interest}
The authors declare that no potential conflicts of interest with respect to the research, authorship, and/or publication of this article.

\section*{Data availability}
Data sharing not applicable to this article as no data sets were generated
or analyzed during the current study.

\renewcommand\baselinestretch{1.3}

\end{document}